\documentclass[prb,superscriptaddress,amsmath,amssymb,twocolumn]{revtex4-2}
\usepackage{natbib}
\usepackage{graphicx}         

\usepackage{textcmds}
\usepackage{dcolumn}          
\usepackage{bm}               
\usepackage{upgreek}
\usepackage{booktabs} 				
\usepackage{tabularx}
\usepackage{array}
\usepackage[colorlinks=true,urlcolor=blue,breaklinks=true]{hyperref}
\usepackage{xcolor}
\usepackage[a4paper,left=2cm,right=2cm]{geometry}

\begin{document}

\title{Single in situ Interface Characterization Composed of Niobium and a Selectively Grown (Bi$_{1-x}$Sb$_x$)$_2$Te$_3$ Topological Insulator Nanoribbon}

\author{Kevin Janßen}
 \affiliation{Peter Gr\"unberg Insitute 9, Forschungszentrum J\"ulich GmbH, 52425 J\"ulich, Germany}%
 \affiliation{Peter Gr\"unberg Insitute 6, Forschungszentrum J\"ulich GmbH, 52425 J\"ulich, Germany}%
 \affiliation{JARA-Fundamentals of Future Information Technology, J\"ulich-Aachen Research Alliance, Forschungszentrum J\"ulich GmbH and RWTH Aachen University, Germany}%

\author{Philipp R\"u{\ss}mann}
\affiliation{Institute for Theoretical Physics and Astrophysics, University of Würzburg, D-97074 Würzburg, Germany}
\affiliation{Peter Gr\"unberg Institute 1 and Institute for Advanced Simulation 1, Forschungszentrum J\"ulich and JARA, 52425 J\"ulich Germany}

\author{Sergej Liberda}
 \affiliation{Peter Gr\"unberg Insitute 9, Forschungszentrum J\"ulich GmbH, 52425 J\"ulich, Germany}%
 \affiliation{JARA-Fundamentals of Future Information Technology, J\"ulich-Aachen Research Alliance, Forschungszentrum J\"ulich GmbH and RWTH Aachen University, Germany}%

\author{Michael Schleenvoigt}
 \affiliation{Peter Gr\"unberg Insitute 9, Forschungszentrum J\"ulich GmbH, 52425 J\"ulich, Germany}%
 \affiliation{JARA-Fundamentals of Future Information Technology, J\"ulich-Aachen Research Alliance, Forschungszentrum J\"ulich GmbH and RWTH Aachen University, Germany}%
 
\author{Xiao Hou}
 \affiliation{Peter Gr\"unberg Insitute 6, Forschungszentrum J\"ulich GmbH, 52425 J\"ulich, Germany}%
 
\author{Abdur Rehman Jalil}
 \affiliation{Peter Gr\"unberg Insitute 9, Forschungszentrum J\"ulich GmbH, 52425 J\"ulich, Germany}%
 \affiliation{JARA-Fundamentals of Future Information Technology, J\"ulich-Aachen Research Alliance, Forschungszentrum J\"ulich GmbH and RWTH Aachen University, Germany}%

\author{Florian Lentz}
 \affiliation{Helmholtz Nano Facility, Forschungszentrum J\"ulich GmbH, 52425 J\"ulich, Germany}
 
\author{Stefan Trellenkamp}
 \affiliation{Helmholtz Nano Facility, Forschungszentrum J\"ulich GmbH, 52425 J\"ulich, Germany}
 
\author{Benjamin Bennemann}
 \affiliation{Peter Gr\"unberg Insitute 9, Forschungszentrum J\"ulich GmbH, 52425 J\"ulich, Germany}%
 \affiliation{JARA-Fundamentals of Future Information Technology, J\"ulich-Aachen Research Alliance, Forschungszentrum J\"ulich GmbH and RWTH Aachen University, Germany}%
 
 \author{Erik Zimmermann}
 \affiliation{Peter Gr\"unberg Insitute 9, Forschungszentrum J\"ulich GmbH, 52425 J\"ulich, Germany}%
 \affiliation{JARA-Fundamentals of Future Information Technology, J\"ulich-Aachen Research Alliance, Forschungszentrum J\"ulich GmbH and RWTH Aachen University, Germany}%
 
\author{Gregor Mussler}
 \affiliation{Peter Gr\"unberg Insitute 9, Forschungszentrum J\"ulich GmbH, 52425 J\"ulich, Germany}%
 \affiliation{JARA-Fundamentals of Future Information Technology, J\"ulich-Aachen Research Alliance, Forschungszentrum J\"ulich GmbH and RWTH Aachen University, Germany}%

\author{Peter Sch\"uffelgen}
 \affiliation{Peter Gr\"unberg Insitute 9, Forschungszentrum J\"ulich GmbH, 52425 J\"ulich, Germany}%
 
 \affiliation{JARA-Fundamentals of Future Information Technology, J\"ulich-Aachen Research Alliance, Forschungszentrum J\"ulich GmbH and RWTH Aachen University, Germany}%
 
\author{Claus-Michael Schneider}
 \affiliation{Peter Gr\"unberg Insitute 6, Forschungszentrum J\"ulich GmbH, 52425 J\"ulich, Germany}%
 \author{Stefan Bl\"ugel}
 \affiliation{Peter Gr\"unberg Institute 1 and Institute for Advanced Simulation 1, Forschungszentrum J\"ulich and JARA, 52425 J\"ulich Germany}
 
\author{Detlev Gr\"utzmacher}
 \affiliation{Peter Gr\"unberg Insitute 9, Forschungszentrum J\"ulich GmbH, 52425 J\"ulich, Germany}%
 \affiliation{JARA-Fundamentals of Future Information Technology, J\"ulich-Aachen Research Alliance, Forschungszentrum J\"ulich GmbH and RWTH Aachen University, Germany}%

\author{Lukasz Plucinski}
 \affiliation{Peter Gr\"unberg Insitute 6, Forschungszentrum J\"ulich GmbH, 52425 J\"ulich, Germany}%
 
\author{Thomas Schäpers}
  \affiliation{Peter Gr\"unberg Insitute 9, Forschungszentrum J\"ulich GmbH, 52425 J\"ulich, Germany}%
 \affiliation{JARA-Fundamentals of Future Information Technology, J\"ulich-Aachen Research Alliance, Forschungszentrum J\"ulich GmbH and RWTH Aachen University, Germany}%
 \email{t.schaepers@fz-juelich.de}

\date{\today}

\begin{abstract}
With increasing attention in Majorana physics for possible quantum bit applications, a large interest has been developed to understand the properties of the interface between a $s$-type superconductor and a topological insulator. Up to this point the interface analysis was mainly focused on in situ prepared Josephson junctions, which consist of two coupled single interfaces or to ex-situ fabricated single interface devices. In our work we utilize a novel fabrication process, combining selective area growth and shadow evaporation which allows the characterization of a single in situ fabricated Nb/$\mathrm{(Bi_{0.15}Sb_{0.85})_2Te_3}$ nano interface. The resulting high interface transparency is apparent by a zero bias conductance increase by a factor of 1.7. Furthermore, we present a comprehensive differential conductance analysis of our single in situ interface for various magnetic fields, temperatures and gate voltages. Additionally, density functional theory calculations of the superconductor/topological insulator interface are performed in order to explain the peak-like shape of our differential conductance spectra and the origin of the observed smearing of conductance features. 
\end{abstract}

\maketitle

\section{Introduction}

With the prediction of Majorana zero modes in $p$-wave superconducting systems, a large interest has grown on the interface physics of three-dimensional topological insulators (3D TIs) and $s$-wave superconductors \cite{wilczek_majorana_2009,fu_superconducting_2008}. Recent studies tried to establish $p$-wave superconductivity by either proximizing or doping a 3D TI with an $s$-wave superconductor  \cite{sacepe_gate-tuned_2011,williams_unconventional_2012,veldhorst_josephson_2012,cho_symmetry_2013,kurter_dynamical_2014,stehno_signature_2016,kurter_evidence_2015,yang_proximity_2012,wiedenmann_transport_2017,kurter_conductance_2019}. Due to the superconducting proximity effect, $p$-wave superconductivity is expected to be established in the interface region and give rise to Majorana zero modes. For the realization of Majorana zero modes, a pristine interface between the two materials is of great importance, since surface oxidation can lead to non-topological states \cite{ngabonziza_situ_2015,thomas_surface_2016}. Currently, the major platform for the analysis of these interfaces is the topological Josephson junction, \cite{mourik_signatures_2012,williams_unconventional_2012,veldhorst_josephson_2012,cho_symmetry_2013,kurter_dynamical_2014,stehno_signature_2016} where 4$\pi$-contributions in the Josephson current are predicted to be a proof of established $p$-wave superconductivity \cite{fu_josephson_2009}. 

Nevertheless, in a Josephson junctions, always two coupled superconductor/topological insulator interfaces are investigated. Such a coupled system gives rise to multiple Andreev reflections and leads to additional effects like a change of the density of states in the junction \cite{octavio_subharmonic_1983}. Furthermore, evidence for a topological proximity effect has been found in angle-resolved photo-emission experiments \cite{shoman_topological_2015,trang_conversion_2020}, which shows that a comprehensive understanding of the interface physics between these two material classes is not established, yet. Therefore, measurements on single interfaces are needed to obtain an in-depth understanding of the interface physics between superconductors and topological insulators. Up to now, such single interfaces have been measured on a variety of systems, like exfoliated flakes \cite{banerjee_signatures_2018} or grown crystals \cite{stehno_conduction_2017, finck_phase_2014, Zhang2021}, but were limited to two-point configurations and ex-situ lift-off processes. However, ex-situ fabrication can lead to degradation of the interface region due to oxidation or surface roughening by Ar milling.

Here, we present a new fabrication process based on selective-area molecular beam epitaxy growt in combination with shadow-mask evaporation. This process enabled us to fabricate and analyze a clean in situ processed single superconductor/topological insulator interface made out of Nb as a superconductor and $\mathrm{(Bi_{0.15}Sb_{0.85})_2Te_3}$ (BST) as a topological insulator. Furthermore, we investigated the response of the differential conductance of the interface on varying temperature, magnetic field and gate voltage. For the theoretical description of the transport, density functional calculations of the superconductor/topological insulator interface were performed, in order explain the origin of the measured conductance features.

\section{Results and Discussion} \label{Res}

\subsection{Material Characterization} \label{sec:Charact}

For the characterization of our $\mathrm{(Bi_{0.15}Sb_{0.85})_2Te_3}$ topological insulator layer, we conducted Hall measurements. Thereby, the resistivity, two-dimensional carrier concentration, and charge carrier mobility were determined to be $\rho = 4.94 \cdot 10^{-4}\, \mathrm{\Omega cm} $, $n_{\mathrm{2D}} = 4.44 \cdot 10^{13} \mathrm{cm}^{-2}$, and $\mu = 285\, \mathrm{cm{^2}/Vs}$ (see Supplementary Information I). Furthermore, the critical temperature and magnetic field of our Nb layer were determined to be $T_{\mathrm{c}} = 7.0\,$K and $B_{\mathrm{c}} = 3.5\,$T, respectively, and were associated with a superconducting band gap of $\Delta_{\mathrm{Nb}} = 1.06\,$meV, following BCS theory (see Supplementary Information II). 

In order to estimate the resistance contribution of the $\mathrm{(Bi_{0.15}Sb_{0.85})_2Te_3}$ ribbon up to the normal contact, a number of reference devices and their interfaces for several distances have been measured. The $\mathrm{(Bi_{0.15}Sb_{0.85})_2Te_3}$ and the Nb film of these reference devices were processed in the exact same run as the presented device. Measurements at DC zero bias for different distances to the superconducting interface lead to an interface resistance at zero bias voltage of $R_{\mathrm{0}} = (200\, \pm \,200)\, \Omega$ and a resistance per length of the 100$\,$nm  $\mathrm{(Bi_{0.15}Sb_{0.85})_2Te_3}$ ribbon of $R/l_{\mathrm{BST}} = (4.9 \pm 0.3)\, \mathrm{k \Omega / \mu m}$ (see linear regression in Supplementary Information III). 

\subsection{Differential Conductance Spectra} \label{sec:Measurement}

\begin{figure}[hbpt]
\includegraphics[width=0.94\columnwidth]{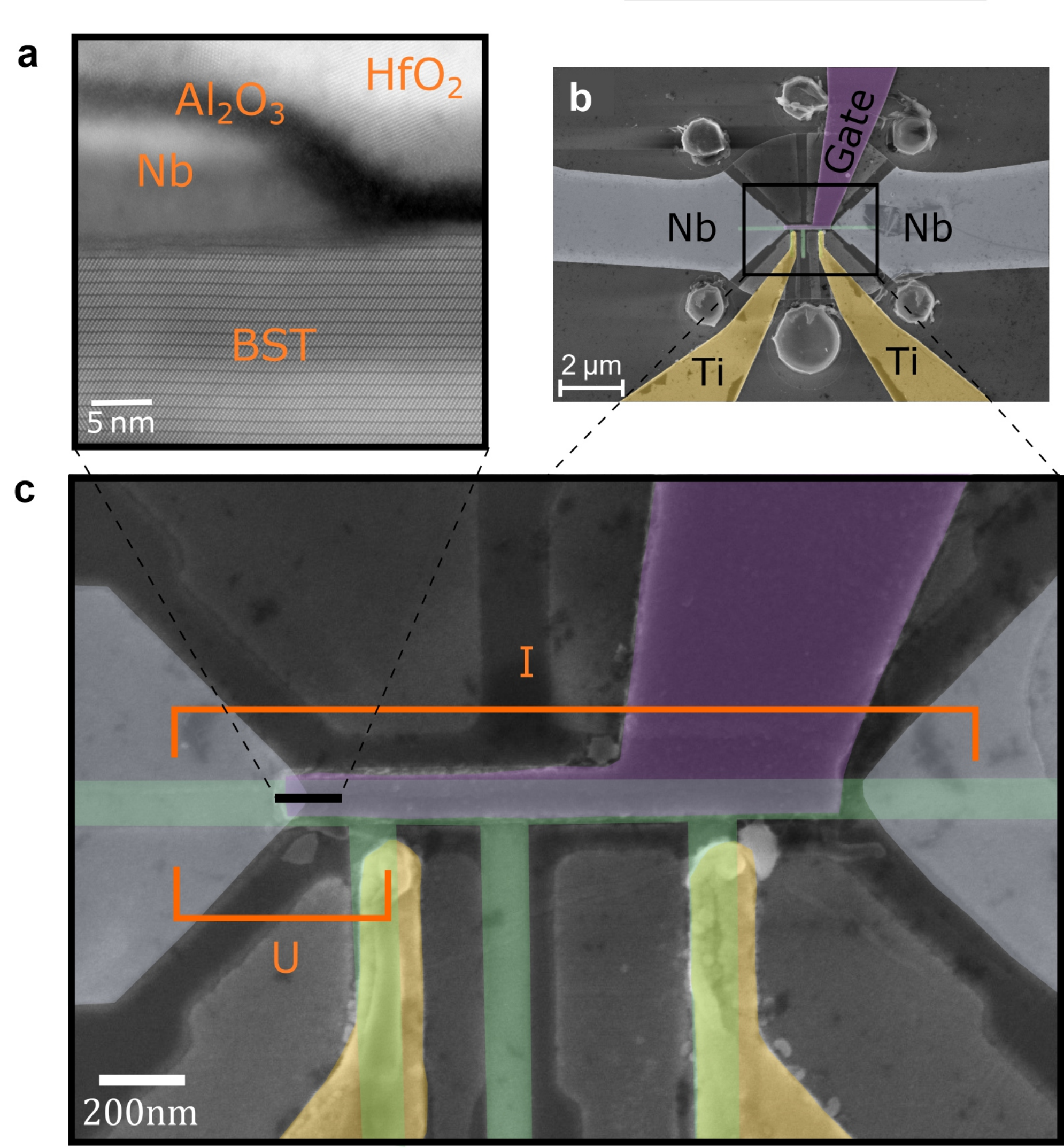}
\caption{\label{fig:Fig0}\textbf{(a)} High-angle annular dark-field scanning transmission electron microscopy image of the interface region. \textbf{(b)} Scanning electron micrograph of the measured device with additional top-gate contact, corresponding to (a)(iv). The Nb contacts are highlighted in light gray, the $\mathrm{(Bi_{0.15}Sb_{0.85})_2Te_3}$ in green, the Ti normal contacts in yellow and the top-gate in purple. \textbf{(c)} Scanning electron micrograph of the device. The black line indicates the region shown in \textbf{(b)} while the orange lines show the contacts where current is applied and voltage is measured. Remaining Nb islands due to stencil mask process are not highlighted with any color since they are not relevant for the transport measurements.}
\end{figure}

Next we analyze the transport across a single $\mathrm{(Bi_{0.15}Sb_{0.85})_2Te_3}$/Nb interface in detail. The corresponding device and the cross section of the interface are depicted in Fig.~\ref{fig:Fig0} (a)-(c). Details on the sample fabrication can be found in the Methods section. 

\begin{figure}[t]
\includegraphics[width=\columnwidth]{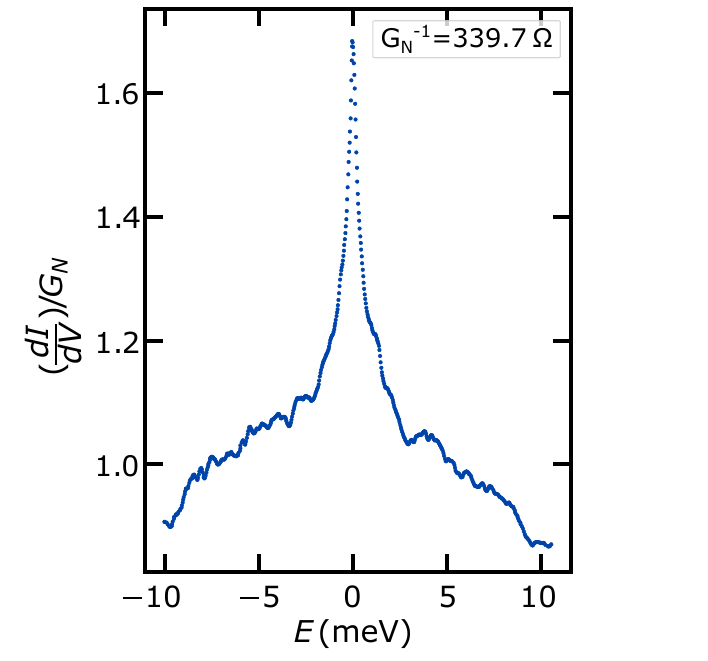}
\caption{\label{fig:Fig1} Normalized differential conductance data (blue) of the Nb/BST interface $(dI/dV)/G_N$ for $R_{\mathrm{BST}} = 1220\, \Omega$ and $G_{\mathrm{N}}^{-1} = 339.7\, \Omega$ at $T = 1.5\,$K.}
\end{figure}

The normalized differential conductance $(dI/dV)/G_N$ as a function energy shown in Fig.~\ref{fig:Fig1} for $T$=1.5~K and zero gate voltage. For this sample the distance between the Nb electrode and the normal contact was 220\,nm. In order to gain detailed information on the interface properties itself, first the resistance contribution $R_{\mathrm{BST}}$ of the $\mathrm{(Bi_{0.15}Sb_{0.85})_2Te_3}$ ribbon up to the normal contact was subtracted. Assuming an interface resistance of $R_{\mathrm{0}}=200\,\Omega$ at zero bias, as given above, we estimate this contribution to be $R_{\mathrm{BST}} = 1220\, \Omega$. The voltage drop at the interface itself was determined by subtracting the according voltage drop in the nanoribbon segment from the total voltage (for the raw data, see Supplementary Information IV).
The normal state conductance $G_{\mathrm{N}}$ is determined by the mean conductance at $\pm 6 \, \Delta_\mathrm{Nb}$. The high interface quality from our in situ process becomes apparent in the large conductance increase around zero bias voltage, as expected for a high contribution of Andreev reflection. This agrees well with the results of Schüffelgen \textit{et al.}~\cite{schuffelgen_selective_2019}, who derived an interface transparency of 0.95, on a Josephson junction device, fabricated with the same in situ approach.

Concerning the Blonder-Tinkham-Klapwijk (BTK) model, such a conductance increase is generally expected, but the sharp and peak-like form does not agree with the model. For the original BTK model the increase is expected to start close to $\Delta$ and not significantly before. A broadening could be achieved with the consideration of inelastic scattering, as presented by Plecen\'ik \textit{et al.} \cite{plecenik_finite-quasiparticle-lifetime_1994}. However, the overall peak-like shape can still not be obtained. Furthermore, our temperature dependent data can exclude a critical current based origin, since the peak height and shape is stable up to 4.5\,K, as discussed later. Additionally, the high $p$-doping of our ribbon results in a low contribution of the topological surface states to the overall current. Therefore, we do not expect that a modeling extension in this manner would be reasonable. However, the BTK model neglects the superconducting proximity effect, which is expected to be especially distinct in high quality interfaces. Therefore, a density functional theory analysis of our interface has been carried out in order to investigate the interface coupling between the TI ribbon and the Nb  (see Sec.~\ref{sec:DFT}).\\
In order to estimate the error of the peak height, we additionally modelled our data for an upper and a lower limiting case of the interface resistance. The upper case was chosen by the error of the linear regression to be $R^{\mathrm{up}}_{\mathrm{0}} = 400\,\Omega$. In order to not exceed the maximal value of $2G_{\mathrm{N}}$ for ideal Andreev reflection, the lower value was limited to $R^{\mathrm{low}}_{\mathrm{0}} = 146\,\Omega$. This procedure results in a maximum peak height of 2 and a minimum peak height of 1.3. In any case a significant zero bias conductance increase is justified.

\subsection{Temperature Dependence}\label{sec:Temp}

\begin{figure*}[t]
\includegraphics[width=17cm]{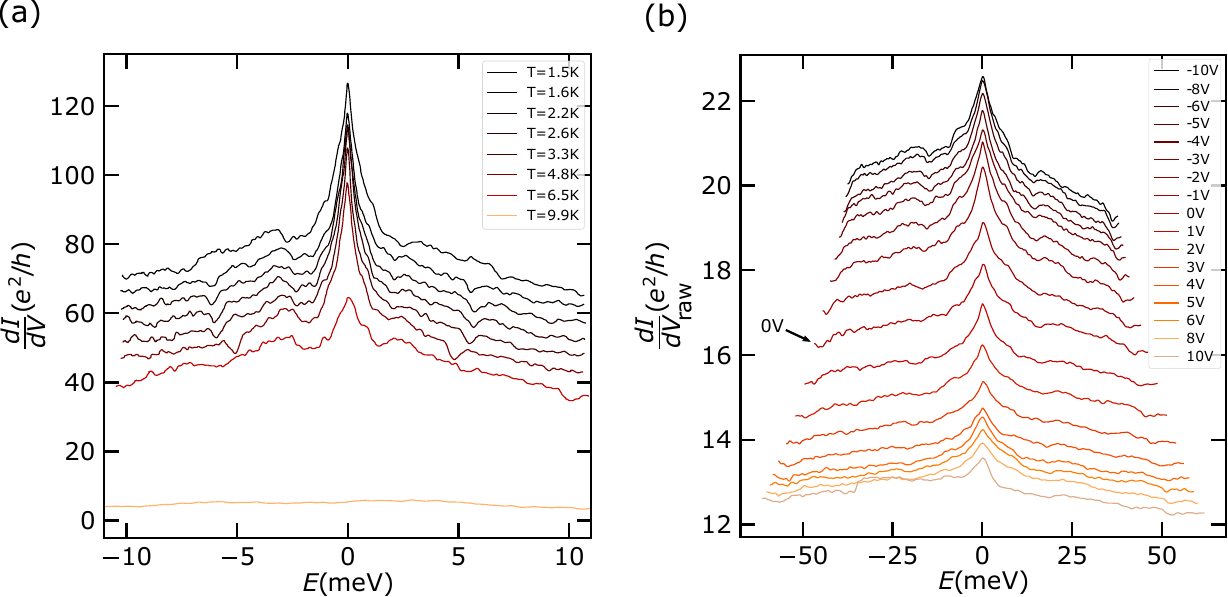}
\caption{\label{fig:Fig2_T}\textbf{(a)} Differential interface conductance for the Nb electrodes in the superconducting state, i.e. temperatures up to 6.5$\,$K and in the normal conducting state at 9.9$\,$K. The curves are offset by $5\,\mathrm{e^2/h}$. \textbf{(b)} Differential conductance without subtraction of $R_{\mathrm{BST}}$ for different gate voltages. }
\end{figure*}

The temperature dependence of the interface properties has been measured for temperatures up to 10$\,$K. As done for the measurement shown in the previous section, the additional resistance of $R_{\mathrm{BST}} = 1220\, \Omega$ has been subtracted (for the raw data, see Supplementary Information VI). The differential conductance as a function of energy is shown in Fig.~\ref{fig:Fig2_T}(a). 
A prominent conductance peak is observed at zero energy with basically no change up to 4.5$\,$K. At a temperature of 
$6.5$\,K the peak height is significantly reduced, whereas at around 10$\,$K, which is already above the superconducting transition temperature of our Nb, no features are observed anymore. This confirms that the zero bias conductance peak originates from the superconducting Nb electrodes. 

With the robustness of the peak height and width up to temperatures of 4.5\,K, which corresponds to half of the superconducting band gap of the Nb film, we can exclude a spurious zero bias conductance, as discussed in \cite{sasaki_topological_2011}. Such a spurious zero bias peak can arise when local currents exceed the critical current and can also lead to a voltage dependent decrease of the differential conductance.

\subsection{Gate Dependence}\label{sec:Topg}

The effect of the top gate on the differential resistance without subtraction of $R_{\mathrm{BST}}$ is shown in  Fig.~\ref{fig:Fig2_T}(b). The gating behaviour of our $\mathrm{(Bi_{0.15}Sb_{0.85})_2Te_3}$ ribbon reveals $p$-type conductance of our composition, since the conductance at zero bias increases with more negative gate voltage.
This agrees well with the findings of  Weyrich \textit{et al.}~\cite{weyrich_growth_2016} who estimated the transition from $n$-type to $p$-type doping for $\mathrm{(Bi_{1-x}Sb_{x})_2Te_3}$ in electron transport for $x = 0.30-0.49$. 
For the spectra no systematic behaviour with gate voltages up to $\pm 10$\,V is observed. Note, the increasing peak height with decreasing gate voltage results from the general increase of conductance by the gate voltage. When subtracting the influence of the gate voltage, no systematic behavior of the peak with gate voltage is observed (see Supplementary Information V). Since, the gate contact is directly above the $\mathrm{(Bi_{0.15}Sb_{0.85})_2Te_3}$/Nb interface region, gating of the interface region and the nanoribbon is assured. Due to the expected screening of the superconducting Nb above the interface region, it is plausible that the interface properties do not change with gate voltage. Since, a mismatch in Fermi velocities between Nb and the TI material contributes to the effective barrier strength, this also indicates that both Fermi velocities are not changed significantly. In case of our $p$-doped $\mathrm{(Bi_{0.15}Sb_{0.85})_2Te_3}$ ribbon this is reasonable, since the Fermi level is expected to be located in the valence band which has a high density of states.

\subsection{Magnetic Field Dependence}\label{sec:Mag}

\begin{figure*}[htbp]
\includegraphics[width=17cm]{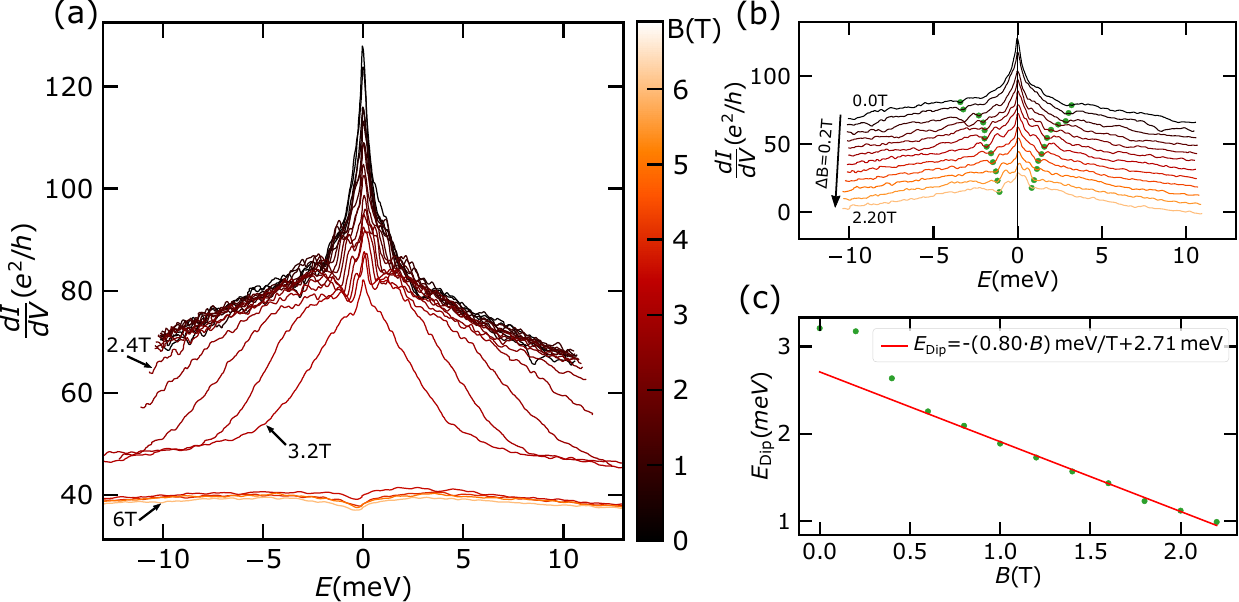}
\caption{\label{fig:Fig3_B}\textbf{(a)} Differential interface conductance for different magnetic fields. \textbf{(b)} Waterfall plot of the differential interface conductance for magnetic fields up to $2.2\,$T. The curves are offset by $6\,\mathrm{e^2/h}$. \textbf{(c)} Position of the conductance dips, highlighted in \textbf{(b)} by green dots, for different magnetic fields.}
\end{figure*}

The out-of-plane magnetic field dependence has been investigated for magnetic fields up to 6$\,$T. The differential conductance of the interface is shown in Fig.~\ref{fig:Fig3_B}(a). As before, a resistance of $1220\,\Omega$ has been attributed to the BST ribbon segment up to the normal contacts and has been subtracted to achieve the resistance of the interface itself. With increasing magnetic field we find a suppression of the zero bias conductance peak. 
From 2.4$\,$T to 3.2$\,$T we observe a region which is driven out of superconductivity for increasing bias voltage. For magnetic fields above $3.6\,$T the superconductivity of our Nb electrodes is completely suppressed, as expected ($B_c = 3.5\,$T, see Supplementary Information II) and no conductance increase for low bias voltages is observed anymore. In contrast to that, a zero bias conductance dip is now observed in our measurements at 4-6$\,$T. We have strong indications that this is an electron-electron interaction based phenomenon originating from the $\mathrm{(Bi_{0.15}Sb_{0.85})_2Te_3}$ film, as previously reported by Stehno \textit{et al.}~\cite{stehno_conduction_2017, tikhonov_andreev_2016}. Additional investigations reveal that the dip is also present in a four terminal measurement only containing the $\mathrm{(Bi_{0.15}Sb_{0.85})_2Te_3}$, see Supplementary Information VII. This allows us to rule out the re-entrance effect as a possible explanation, as reported for the data of Finck \textit{et al.} \cite{finck_phase_2014}. The dip is stable with magnetic field up to 6$\,$T and vanishes with increasing temperature, since its not present at 9.9$\,$K, see Fig.~\ref{fig:Fig2_T}(a). This behaviour agrees with the findings of Stehno \textit{et al.}~\cite{stehno_conduction_2017}, who attributed the dip to electron-electron interaction.\\

Figure~\ref{fig:Fig3_B}(b) shows the differential conductance up to 2.2$\,$T in a waterfall plot. We observe conductance dips which are marked by green dots. From 0.5$\,$T the magnetic field response of the dip position can be described by a linear relation, as shown in Fig.~\ref{fig:Fig3_B}(c). The position is clearly outside the superconducting band gap of our Nb layer and even a second set of dips with a similar behaviour, starting at 9$\,$meV for zero magnetic field can be observed. Such dips have been observed before with different possible explanations \cite{banerjee_signatures_2018,voerman_dominant_2019,li_origin_2017}. Because of the response to magnetic field and the high voltage at which the dips occur, we expect these features to originate from the bulk Nb contact and to not be related to the interface region.

\subsection{Superconducting density of states of the superconductor-TI interface from density functional theory}\label{sec:DFT}

\begin{figure}[htbp]
    \centering
    \includegraphics[width=8cm]{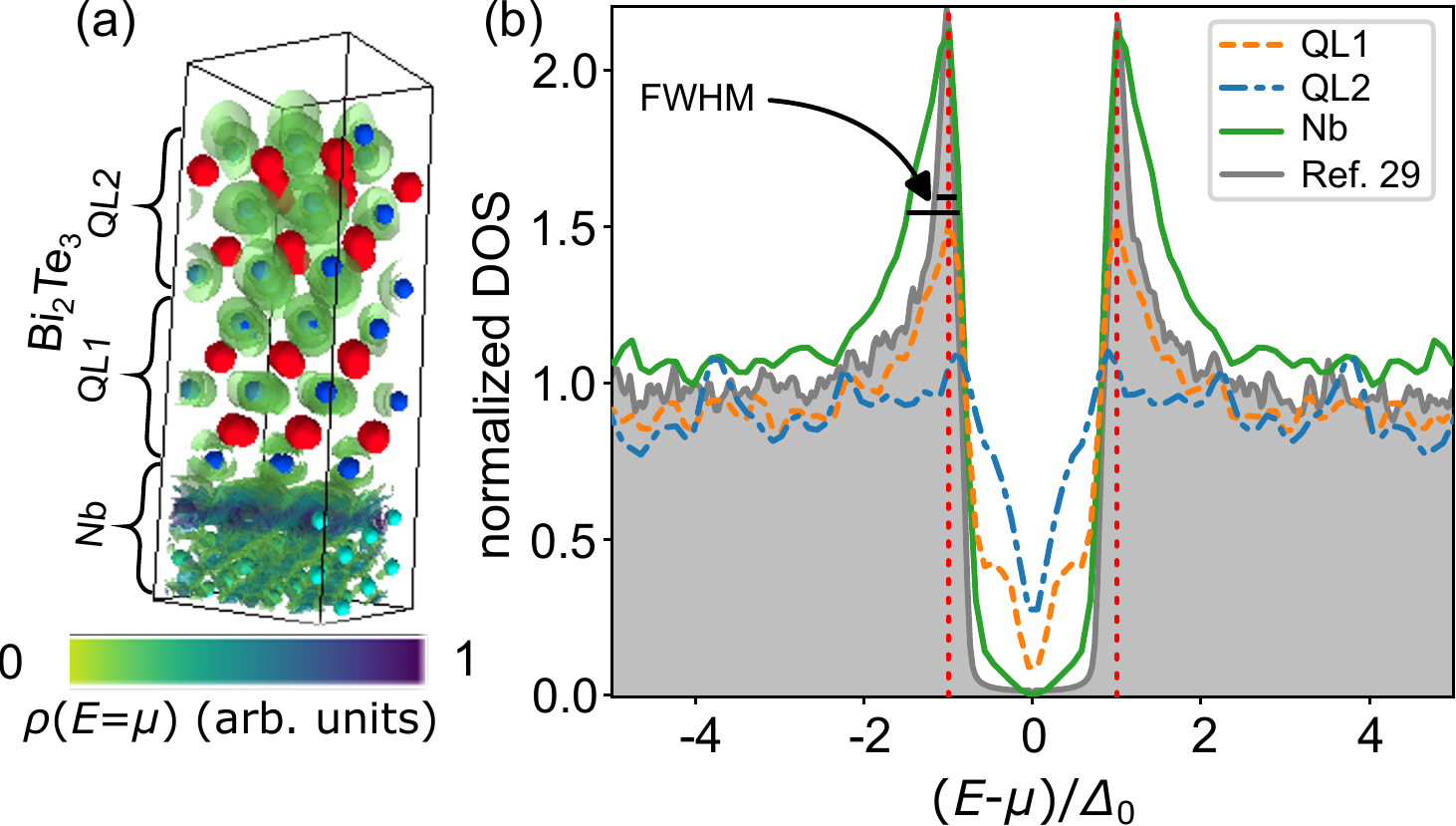}
    \caption{\label{fig:Fig6}\textbf{(a)} DFT-calculated contribution from the Fermi energy to the charge density in the superconductor/TI heterostructures for $p$-type Bi$_2$Te$_3$. \textbf{(b)} Superconducting density of states in the superconductor (Nb) and the first and second quintuple layers of the TI.  The grey background shows the superconducting DOS for a clean Nb surface~\cite{rusmann_density_2022}. The black bars indicates the full width at half maximum (FWHM) of the Nb coherence peaks.}\end{figure}

In this section we will discuss the origin of the shape of our conductance peak and the smearing in the conductance dips. To this end, we conducted state-of-the-art DFT calculations of the density of states in the interface region for various chemical potentials $\mu$. We employ the Kohn-Sham Bogoliubov-de Gennes method taking into account all details of the electronic structure to model superconductivity~\cite{rusmann_density_2022}, which is described in Sec.~\ref{sec:MethodDFT}. Our DFT results for the superconductor-topological insulator (SC-TI) interface are summarized in Fig.~\ref{fig:Fig6}. With varying position of the chemical potential $\mu$ in the TI, the density of states at the chemical potential $\rho(E=\mu)$ changes drastically when $\mu$ lies in the bulk valence band (VB) or conduction band (CB) of the TI compared to the case when it resides inside the bulk band gap (see also Supplementary Section IX). In the latter case, only the topological surface state contributes to $\rho(E=\mu)$ and very few states are available for hybridization with the electronic structure of the superconductor. 
This is seen in the contribution to the normal state density at $\mu$, integrated over the TI region, which is reported in Tab.~\ref{tab:FWHM_DFT}. This fact is illustrated in Fig.~\ref{fig:Fig6}(a) where the charge density at the chemical potential is visualized throughout the SC-TI heterostructure for a $p$-type TI. 

Figure~\ref{fig:Fig6}(b) shows the corresponding superconducting density of states (DOS), integrated over (i) the superconductor, (ii) the first, and (iii) the second quintuple layer (QL) of the TI in the SC-TI heterostructure. When compared to the coherence peak of a clean Nb surface~\cite{rusmann_density_2022}, we find that the increased hybridization with states of bulk-conducting TI leads to a distinct broadening of the superconducting coherence peak by more than a factor two. We measure this with the full width at half maximum (FWHM) of the coherence peak that is given in Tab.~\ref{tab:FWHM_DFT}. We stress that the coherence peak mainly originates from the Nb layers in the simulation and is strongly suppressed in the local DOS within the TI, where it decays with the distance from the surface. Thus, the broadening of the coherence peak is a result of the inverse proximity effect~\cite{Deutscher_deGennes_1969}.

Additionally, stronger in-gap features are visible in the DOS which further washes out the sharp coherence peaks coming from Nb. We attribute the in-gap features to the reduced gap size in the TI which are only proximity-coupled without any intrinsic superconductivity in the TI.
A comparison of the first to the second QL (QL1/2) reveals the decay of the proximity-induced gap in the TI electronic structure~\cite{Chiu2016} which is evident by the flattening-out of the DOS with larger distance from the SC contact. We expect this flattening of the DOS, which gets exceptionally clear in QL2, to be the origin for the sharp and peak like shape of our differential conductance spectra. It is noteworthy that we observe similar behavior for $\mu$ inside the VB and the CB where in both cases the normal state charge density at $\mu$ in the TI is roughly a factor 2 larger than when $\mu$ lies in the TI gap (see Tab.~\ref{tab:FWHM_DFT}).

Our DFT data for the SC/TI interface with $p$-type Bi$_2$Te$_3$ can be linked to the transport data obtained for our samples. The transport measurements proved the high quality of the interface with a robust zero-bias conductance peak. Our DFT data qualitatively proves that a good electrical coupling, which is the case in the in situ-grown samples of this work, leads to significant hybridization and a proximity effect in the TI. This in turn results in a broadening of the superconducting coherence peaks and a narrowing of the DOS inside the superconducting band gap. According to the BTK model, the transport features are closely connected to the shape of the superconducting DOS. The broadening of the coherence peak thus can be connected to the largely feature-less conductance data. Therefore, we conclude that our DFT data qualitatively explains the origin of the sizable broadening in the zero bias conductance peak.

\begin{table}[hbt]
    \centering
    \caption{\label{tab:FWHM_DFT}Average full width at half maximum $\langle\mathrm{FWHM}\rangle$ (in units of the intrinsic gap size of the superconductor $\Delta_0$) of the coherence peaks for different locations of the chemical potential ($\mu$) in the TI. The FWHM is averaged over the first 5 layers in the Nb superconductor and over the coherence peak at positive and negative energies. The third column shows the normal state DOS integrated in the TI in 1/eV.
}
    \begin{tabular}{c|c|c|c}
         $\mu$ of the TI & $\langle \mathrm{FWHM}\rangle$ & $\int_{V_\mathrm{TI}}\rho(\mathbf{r}, E=\mu) d^3r$ & reference\\\hline
        in gap &  0.39 & 0.49 & \textrm{this work}\\
        in VB & 0.49 & 1.32 & \textrm{this work} \\
        in CB & 0.47 & 0.88 & \textrm{this work} \\
        -- & 0.22 & -- & Ref.~\onlinecite{rusmann_density_2022}
    \end{tabular}
\end{table}

\section{Conclusion}
In summary, we presented a new in situ fabrication scheme for the fabrication of pristine high quality superconductor/topological insulator single interfaces. With the analysis of our differential conductance data we can confirm the high interface quality with a resulting zero bias conductance increase by a factor of almost 1.7. Furthermore, we deliver a comprehensive analysis of the interface including studies of perpendicular magnetic field, temperature and top gate dependence. The robustness of the zero bias conductance peak to temperatures up to 4.5\,K excludes a critical current induced origin and confirms that the peak originates from Andreev reflection at the interface. The gate dependent measurements showed that the interface properties do not change significantly with gate voltage. Furthermore, we could reveal signatures of electron-electron interaction in $\mathrm{(Bi_{0.15}Sb_{0.85})_2Te_3}$ and confirm the interpretation of Stehno \textit{et al.}~\cite{stehno_conduction_2017}. Since, our differential conductance spectra are not well represented by the BTK modeling, a DFT-based analysis has been carried out. Our DFT-based analysis provides further evidence for the strong electrical coupling that influences the proximity effect in the SC/TI heterostructure. This results in strong hybridization and a broadening of the superconducting coherence peak at the interface, which is in line with our transport data. Furthermore, a strong decrease of the DOS is observed inside the gap for the first QLs. We expect this decrease to be the reason for the peak-like shape of our differential conductance spectra and the deviation to the Blonder-Tinkham-Klapwik model.
With our investigations on a single in situ interface of superconductor and topological insulator we deliver a solid foundation for understanding the interface physics between these two material classes and contribute to the  research on topological quantum bits.

\section{Methods}\label{sec:Methods}

\subsection{Sample fabrication}\label{sec:Exp}

\begin{figure*}[hbpt]
\includegraphics[width=0.94\textwidth]{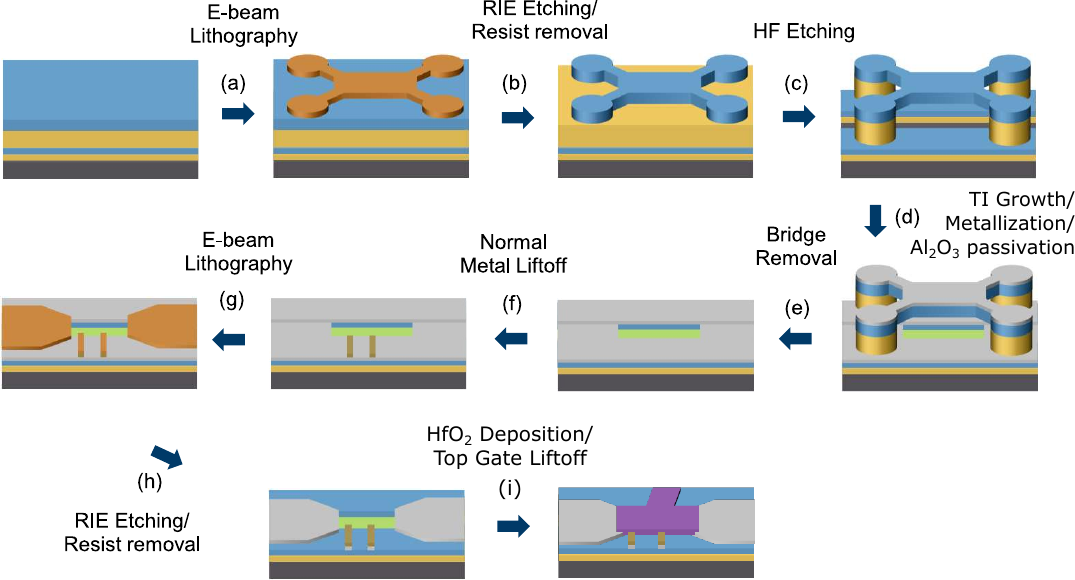}
\caption{Schematic illustration of the fabrication step sequence, with $\mathrm{Si_3N_4}$ in blue, $\mathrm{SiO_2}$ in yellow, Si in black, resist in brown, topological insulator in green, Nb in grey, Ti normal contacts in dark yellow and the Ti top gate in purple.}
\label{fig:Fig0S}
\end{figure*}

For the fabrication of our devices we utilize a combination of selective-area growth and shadow mask evaporation technique \cite{schuffelgen_selective_2019,Schmitt2022,Koelzer2023}. The fabrication step sequence is illustrated in Fig.~\ref{fig:Fig0S}. In order to fabricate our substrates, a 4'' silicon (111) wafer ($\rho >2000\, \Omega \mathrm{cm}$) is first thermally oxidized with 7$\,$nm $\mathrm{SiO_2}$ and then a 30$\,$nm $\mathrm{Si_3N_4}$ film is deposited via plasma-enhanced chemical vapor deposition at $350^\circ$C. In the next step, the trenches for the selective-area growth are etched in the $\mathrm{Si_3N_4}$ layer utilizing electron beam lithography and $\mathrm{CHF_3}$/$\mathrm{O_2}$-based reactive ion etching \cite{jalil2023}. For the stencil mask 300$\,$nm $\mathrm{SiO_2}$ and 100$\,$nm $\mathrm{Si_3N_4}$, are deposited. The stencil mask $\mathrm{Si_3N_4}$ layer is structured by an electron beam lithography process and $\mathrm{CHF_3}$/$\mathrm{O_2}$ reactive ion etching, see Figs.~\ref{fig:Fig0S}(a) and (b), respectively. A 12\% buffered HF etch is utilized to under etch $\mathrm{Si_3N_4}$ to define the stencil bridge structure.
Simultaneously, the the predefined trenches in the lower double layer are revealed as well, as depicted in Fig.~\ref{fig:Fig0S}(c).
Prior to the crystal growth, the sample is etched with 1$\%$ hydrofluoric acid to remove surface oxides. Immediately after, $\mathrm{(Bi_{0.15}Sb_{0.85})_2Te_3}$ is selectively grown via molecular beam epitaxy. The specific composition is chosen such that the Fermi level is close to the Dirac point \cite{weyrich_growth_2016}. In our selective-area growth scheme, the $\mathrm{(Bi_{0.15}Sb_{0.85})_2Te_3}$ is only growing in the predefined Si trenches which expose the Si(111) surfaces. During the crystal growth, the sample is under constant rotation to assure a homogeneous distribution of the components from the different effusion cells and a smooth growth underneath the $\mathrm{Si_3N_4}$ bridge. Subsequently, a 11\,nm thick Nb film is deposited by electron beam evaporation under a fixed angle. During this evaporation the $\mathrm{Si_3N_4}$ bridge is shadowing part of the $\mathrm{(Bi_{0.15}Sb_{0.85})_2Te_3}$ and pre-defines the contacts. For passivation a stoichiometric 3$\,$nm $\mathrm{Al_2O_3}$ layer is deposited on the whole sample under rotation (d).  Afterwards, the shadow mask is removed by mechanical scratching with a cleanroom tissue (e), while the sample is covered with a thin layer of PMMA.

For the fabrication normal contact fingers Ti is deposited by electron beam evaporation and lift-off with a previous Ar-milling step to remove residual surface oxides (f). In the next step, the superconducting contacts are structured by another lithography step (g) and reactive ion etching (h). Prior to the etching, the $\mathrm{Al_2O_3}$ capping on top of the Nb is removed by a dip in 0.2$\%$ hydrofluoric acid. In the last step, the Ti top-gate contact is fabricated (i). Therefore, 16$\,$nm $\mathrm{HfO_2}$ are deposited with atomic layer deposition. Subsequently, the Ti top-gate contact is prepared by a PMMA-based lift-off procedure. Scanning electron micrographs of the device measured in this work are depicted in Figs.~\ref{fig:Fig0}(c) and (d). The 100-nm-wide  selectively-grown $\mathrm{(Bi_{0.15}Sb_{0.85})_2Te_3}$ nanoribbon is contacted on both ends by superconducting Nb electrodes. The two Ti normal contacts are placed at different positions along the nanoribbon. In the high-angular annular dark-field scanning transmission electron microscopy (HAADF-STEM) image shown in Fig.~\ref{fig:Fig0}(a) it can be seen that the interface region between Nb and BST is limited to an intermixing region of a couple quintuple layers (See detailed larger scale image in Supplementary Material VIII) \cite{jalil_engineering_2022}. The atomic resolution gives no indication for a damage of the BST underneath the Nb layer, which confirms the expectations of a clean interface. 

\subsection{Magnetortransport measurements}

Our measurements have been conducted in a He-4 variable temperature insert  setup with a base temperature of 1.5$\,$K and a perpendicular magnetic field of up to 6$\,$T. Thereby, we conducted quasi-DC lock-in amplifier based current driven differential conductance measurements at a frequency of $31.7\,$Hz and with an amplitude of $I_{\mathrm{AC}}=100\,$nA. The bias dependence was investigated by superimposing a DC current in the range $\pm$30$\,\mathrm{\upmu A}$ between both superconducting electrodes. As illustrated in Fig.~\ref{fig:Fig0}(d), the differential conductance was measured between one superconducting contact and one of the normal contacts.

\subsection{Density functional theory}\label{sec:MethodDFT}

In our density functional theory (DFT) calculations we employ the full-potential relativistic Korringa-Kohn-Rostoker Green function method (KKR) \cite{ebert_calculating_2011} as implemented in the JuKKR code \cite{the_jukkr_developers_julich_2022}. Superconducting properties are calculated with the help of the Kohn-Sham-Bogoliubov-de Gennes extension to the JuKKR code \cite{rusmann_density_2022}. We parametrize the normal state exchange correlation functional using the local density approximation (LDA) \cite{vosko_accurate_1980}. We use an $\ell_\mathrm{max}=2$ cutoff in the angular momentum expansion of the space filling Voronoi cells around the atomic centers, where the exact (i.e.\ full-potential) description of the atomic shapes is taken into account \cite{stefanou_efficient_1990,stefanou_calculation_1991}.

The structure we study is an interface between the $s$-wave superconductor Nb and the prototypical TI Bi$_2$Te$_3$, which was previously discussed in Ref.~\onlinecite{rusmann_proximity_2022}. The TI film we use in the calculations for this work consist of 2QL thick Bi$_2$Te$_3$ in contact to nine layers of Nb(111) lattice matched to the in-plane unit cell of the TI. In order to study a shift of the TI Fermi level relative to the electronic structure of the superconductor we employ a renormalization of the energy integration weights during self-consistency in accordance to Lloyd's formula in the KKR method \cite{zeller_elementary_2004}.

The series of DFT calculations in this study are orchestrated with the help of the AiiDA-KKR plugin~\cite{rusmann_aiida-kkr_2021} to the AiiDA infrastructure \cite{huber_aiida_2020}. This has the advantage that the full data provenance (including all values of numerical cutoffs and input parameters to the calculation) is automatically stored in compliance to the FAIR principles of open research data \cite{wilkinson_fair_2016}. The complete data set, that includes the full provenance of the DFT calculations, is made publicly available in the materials cloud archive \cite{talirz_materials_2020,russmann_dataset_2023}. The source codes of the AiiDA-KKR plugin and the JuKKR code are published as open source software under the MIT license~\cite{rusmann_judftteamaiida-kkr_2020,the_jukkr_developers_julich_2022}.

\section{Author Contributions}
K.J., S.L. and A.R.J. designed the devices. K.J. and S.L. fabricated the samples in the clean room. F.L. and S.T. performed all e-beam writing steps. M.S., P.S., G.M. were responsible for TI growth and optimization. B.B. grew the Nb, $\mathrm{Al_2O_3}$ and $\mathrm{HfO_2}$ layers. K.J. performed the electrical measurements of the devices and the analysis. E.Z. helped with the setup of the electrical measurement system. X.H. and A.R.J. performed the HAADF-STEM measurements. P.R. performed the DFT calculations. K.J., P.R., and T.S. wrote the manuscript with contributions of all coauthors. T.S., L.P. supervised the project.

\section{Acknowledgement}

We thank Elmar Neumann for TEM  preparation and Joachim Mayer for feedback on STEM measurements. We thank the board of directors at Forschungszentrum J\"ulich for financial support doctoral projects within the framework of the
strategic further development of method-oriented research. PR and SB acknowledge funding by the Deutsche Forschungsgemeinschaft (DFG, German Research Foundation) under Germany's Excellence Strategy -- Cluster of Excellence Matter and Light for Quantum Computing (ML4Q) EXC 2004/1 -- 390534769 and PR thanks the Bavarian Ministry of Economic Affairs, Regional Development and Energy for financial support within High-Tech Agenda Project ``Bausteine f\"ur das Quantencomputing auf Basis topologischer Materialien mit experimentellen und theoretischen Ans\"atzen''. We are grateful for computing time granted by the JARA Vergabegremium and provided on the JARA Partition part of the supercomputer CLAIX at RWTH Aachen University (project number jara0191).


%

\clearpage
\widetext

\setcounter{section}{0}
\setcounter{equation}{0}
\setcounter{figure}{0}
\setcounter{table}{0}
\setcounter{page}{1}
\makeatletter
\renewcommand{\thesection}{S\Roman{section}}
\renewcommand{\thesubsection}{\Alph{subsection}}
\renewcommand{\theequation}{S\arabic{equation}}
\renewcommand{\thefigure}{S\arabic{figure}}
\renewcommand{\figurename}{Supplementary Figure}
\renewcommand{\bibnumfmt}[1]{[S#1]}
\renewcommand{\citenumfont}[1]{S#1}

\begin{center}
\textbf{Supplementary Material: Single in situ Interface Characterization Composed of Niobium and a Selectively Grown (Bi$_{1-x}$Sb$_x$)$_2$Te$_3$ Topological Insulator Nanoribbon}
\end{center}

\section{Hall measurements of the topological insulator layer}\label{sec:I}

Hall measurements have been conducted on 100\,nm wide reference devices of the exact same molecular beam epitaxy growth and Nb deposition run as the sample presented in the main manuscript. Hall measurements on two reference devices were conducted. The data with corresponding Hall slopes is shown in Supplementary Fig.~\ref{fig:Supp_Hall}.
\begin{figure}[h!]
\includegraphics[width=17cm]{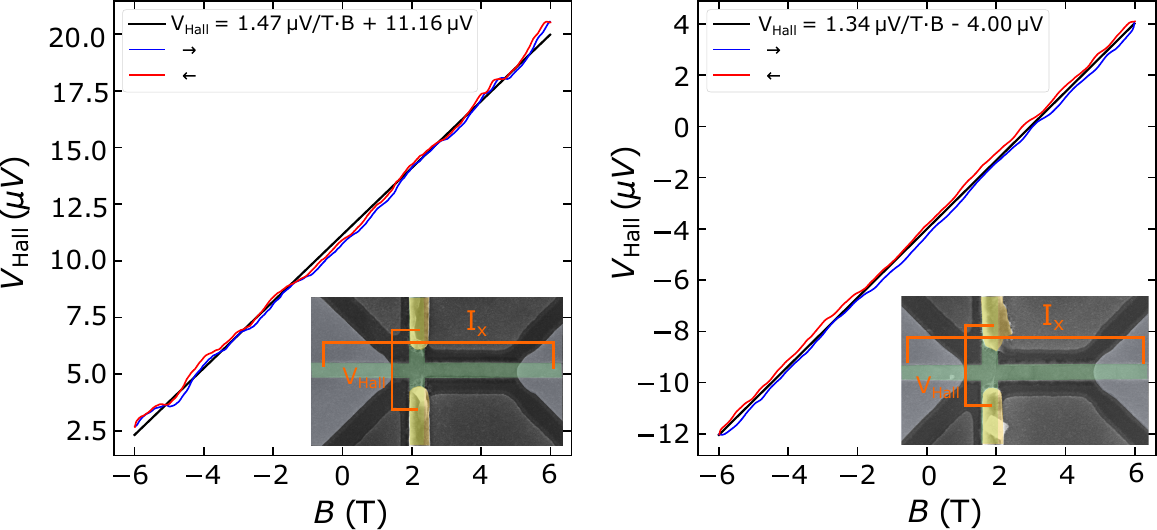}
\caption{\label{fig:Supp_Hall} Hall voltage as a function of magnetic field measured for two reference devices. The colors blue and red distinguish between the forward and backward scan of the magnetic field. Scanning electron micrographs of the corresponding devices and the electrical contacting are shown in the inset. The superconducting contacts are highlighted in grey, the normal contacts in yellow, the $\mathrm{(Bi_{0.15}Sb_{0.85})_2Te_3}$ ribbon in green.}
\end{figure}
\\
The charge carrier density is determined from: 
\begin{equation}
    n_{\mathrm{2D}} = \frac{I_{\mathrm{x}}}{e}\cdot \frac{1}{\frac{dV_{\mathrm{Hall}}}{dB_{\mathrm{z}}}} = 4.44\cdot 10^{13}\, \mathrm{cm}^{-2} \, ,
\end{equation}
with $I_{\mathrm{x}}=100\,$nA the bias current, $e$ the electron charge, and $dV_{\mathrm{Hall}}/dB_{\mathrm{z}}$ the mean slope of both measurements consisting of forward and backward sweep direction. The slope of the Hall curve reveals p-type doping of the $\mathrm{(Bi_{0.15}Sb_{0.85})_2Te_3}$ ribbon. Furthermore, we determined the resistivity $\rho$ from the results of our linear regression presented in Supplementary Sec.~\ref{sec:I}: 
\begin{equation}
    \rho =  a \cdot w \cdot t = 4.94 \cdot 10^{-4} \, \Omega \mathrm{cm} \, , 
\end{equation}
with $a = 4.94\, \mathrm{k\Omega / \upmu m}$ resistance per unit length, $w = 100\,$nm the ribbon width and $t = 10$\,nm the ribbon thickness. The mobility of our $\mathrm{(Bi_{0.15}Sb_{0.85})_2Te_3}$ ribbon can then be estimated by: 
\begin{equation}
    \mu = \frac{t}{\rho \cdot e \cdot  n_{\mathrm{2D}}} = 285 \, \frac{\mathrm{cm}^2}{\mathrm{Vs}} \, . 
\end{equation}

\section{Characterisation of the niobium layer}\label{sec:II} 

\begin{figure}[b!]
\includegraphics[width=\textwidth]{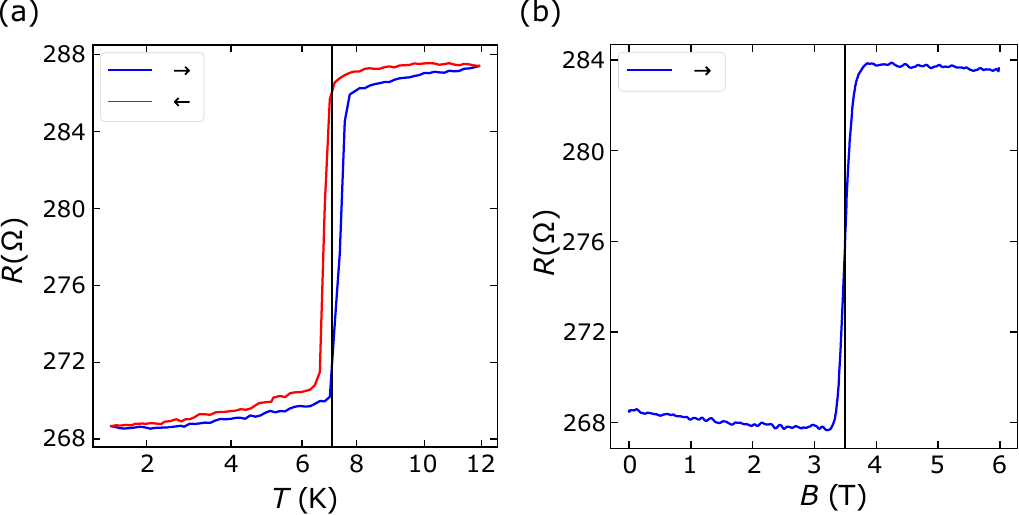}
\caption{\label{fig:Supp_Nb} Resistance of Nb layer measured in a two-point configuration for \textbf{(a)} varying temperature and \textbf{(b)} varying perpendicular magnetic field. The colors blue and red distinguish between the forwards and backwards direction.}
\end{figure}

The thickness of the Nb layer has been determined to 11\,nm and the $\mathrm{Al_2O_3}$ capping to 3\,nm using X-ray-reflectometry. The critical temperature and critical magnetic field were determined to $T_c = 7.0\,$K and $B_c = 3.5\,$T, respectively, by two-point-measurements of the left Nb contact (cf. Supplementary Fig. \ref{fig:Supp_Nb}). 
The offset value of 268\,$\Omega$ is the residual resistance of the cables in the measurement setup.
The measured temperature values were re-calibrated by an additional temperature curve with a Lakeshore  CX-1030-SD-HT-1.4L temperature sensor (Serial number X147729) on the exact sample position. Afterwards, the scaling of the temperature axis was adjusted to these re-calibrated temperature values.
The superconducting gap was determined according to BCS theory:
\begin{equation}
    \Delta_{\mathrm{BCS}} = 1.764 \,k_{\mathrm{B}}  T_{\mathrm{c}} = 1.06\, \mathrm{meV} \; . 
\end{equation}

\section{Resistance per length of topological insulator ribbon }\label{sec:III}

\begin{figure}[h!]
\includegraphics[width=0.45\textwidth]{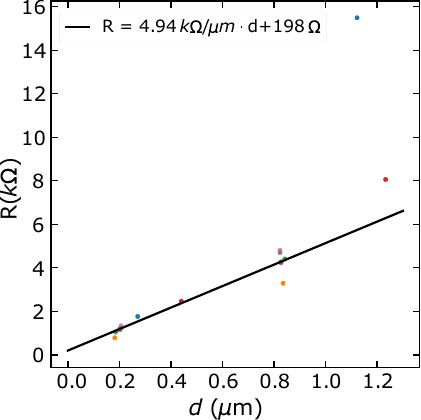}
\caption{\label{fig:Supp_Linfit} Differential resistance $R$ at zero bias voltage for different interfaces and distances $d$. The different colors distinguish between different devices. The black line shows the linear regression of the data, leaving out the two data points for $d$ larger 1$\, \mathrm{\mu m}$.}
\end{figure}

In order to estimate the additional resistance of the $\mathrm{(Bi_{0.15}Sb_{0.85})_2Te_3}$ (BST) nanoribbon from the interface to the superconductor until the normal contact is reached, the resistance has been measured for several distances of the normal contact to the interface using the transmission line method. The measurements were conducted on 100\,nm wide devices on a reference sample. The $\mathrm{(Bi_{0.15}Sb_{0.85})_2Te_3}$ and the Nb on this sample have been processed in the exact same run as the presented device in the main manuscript. Supplementary Fig.~\ref{fig:Supp_Linfit} shows the acquired resistances for different normal contact distances to the superconducting interface. 
Note that for distances above 1$\,\mu$m a low film quality was observed on certain parts of the $\mathrm{(Bi_{0.15}Sb_{0.85})_2Te_3}$ nanoribbon by scanning electron microscopy. The reason for this is the very large $\mathrm{Si_3N_4}$ shadow mask on these particular devices, which impedes a uniform growth. Consequently, this leads to exceptionally high resistance values which do not describe the behaviour of a closed high quality film. Therefore, the two data points for distances above 1$\, \mu $m were left out from the linear regression. The resistance of the nanoribbons segment can be described by: 
\begin{equation}
    R = a \cdot d + b \; , 
\end{equation}
where $a$ is the resistance per unit length and $b$ is the residual resistance, which is the $\mathrm{(Bi_{0.15}Sb_{0.85})_2Te_3}$/Nb interface resistance. For our measurement data we extracted:
\begin{equation}
    a = (4.94 \pm 0.34)\, \mathrm{k}\Omega /\mu \mathrm{m} \, , 
\end{equation}
and
\begin{equation}
    b = (0.20 \pm 0.20)\, \mathrm{k}\Omega \, . 
\end{equation}
Note, that the error on the interface resistance $b$ is equally large to the value itself. Therefore, the value can only be used as an indication when compared to other results.
\clearpage
\section{Raw Data of the differential conductance  spectra\label{sec:IV}}
In Supplementary Fig.~\ref{fig:Supp_dIdVraw} the raw data of the differential conductance at 1.5\,K and zero gate voltage are shown.
\begin{figure}[h!]
\includegraphics[width=8cm]{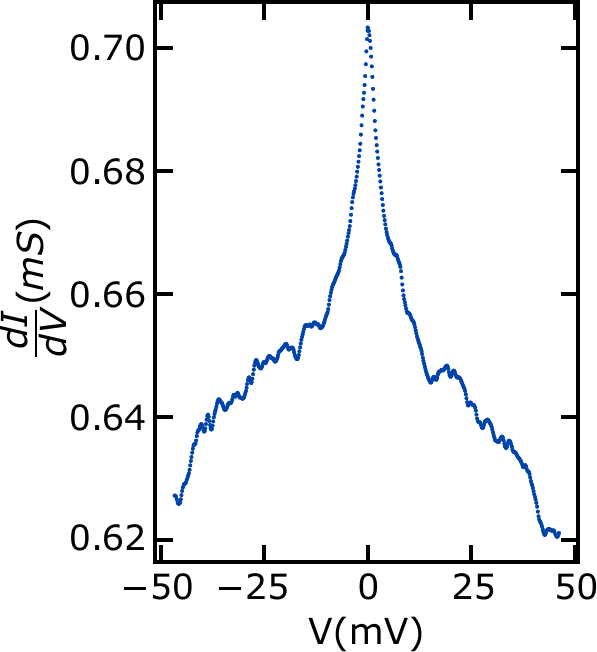}
\caption{\label{fig:Supp_dIdVraw} Raw data of the differential conductance $dI/dV$ for different bias voltages.}
\end{figure}
\clearpage
\section{Gate Dependence of the Differential Conductance}\label{sec:V}
In Supplementary Fig.~\ref{fig:Supp_Fig_Gateincluence} the differential conductance after subtraction of the influence of the gate voltage at zero bias voltage is shown. Therefore, the change in resistance at zero bias voltage has been subtracted from the curves with non-zero gate voltage, before calculating the differential resistance. Fig.~\ref{fig:Supp_Fig_Gateincluence} shows that no systematic behavior with gate voltage is present.
\begin{figure}[htbp]
\includegraphics[width=10cm]{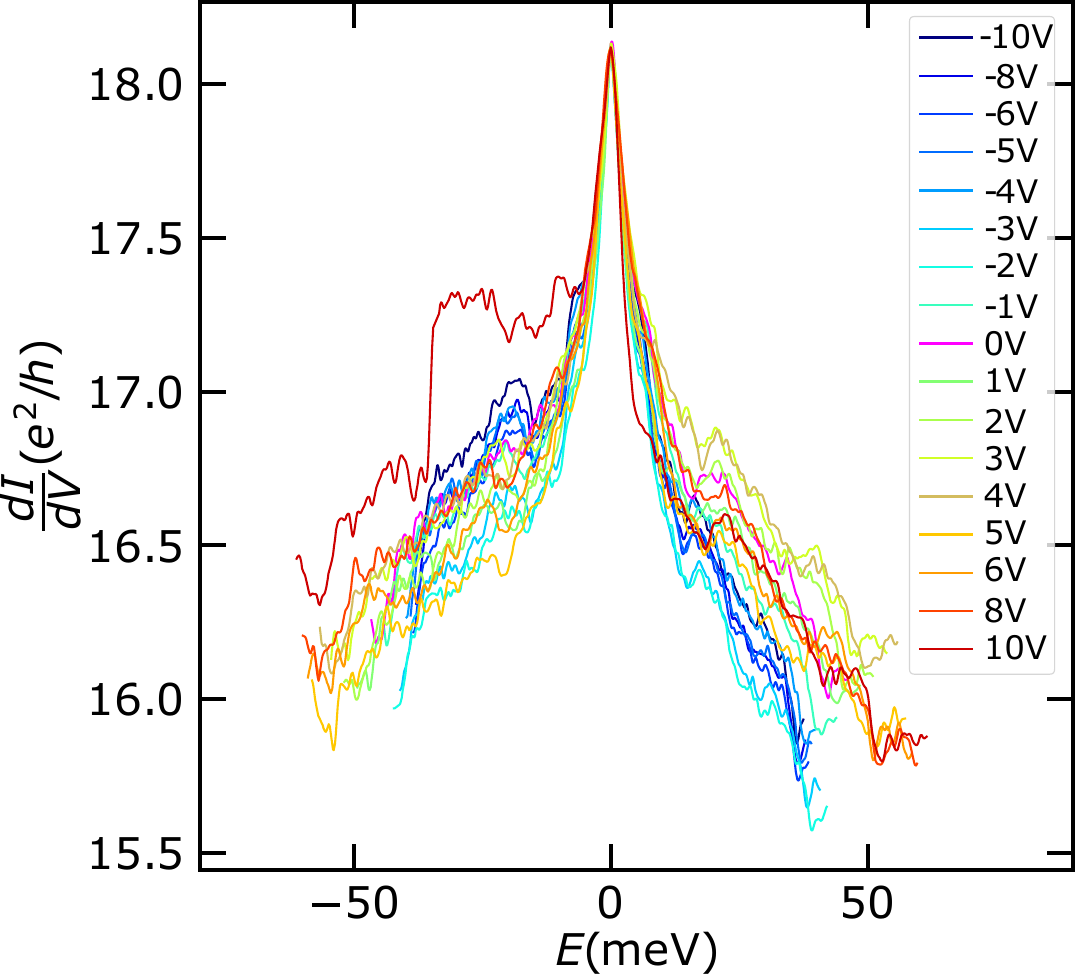}
\caption{\label{fig:Supp_Fig_Gateincluence} Differential conductance after subtracting the influence of the gate voltage at zero bias voltage.}
\end{figure}
\clearpage
\section{Raw Data of the differential conductance for various temperatures} \label{sec:VI}
In Supplementary Fig.~\ref{fig:Supp_Traw} the raw data of the differential conductance for different temperatures at zero gate voltage are shown.
\begin{figure}[h!]
\includegraphics[width=12cm]{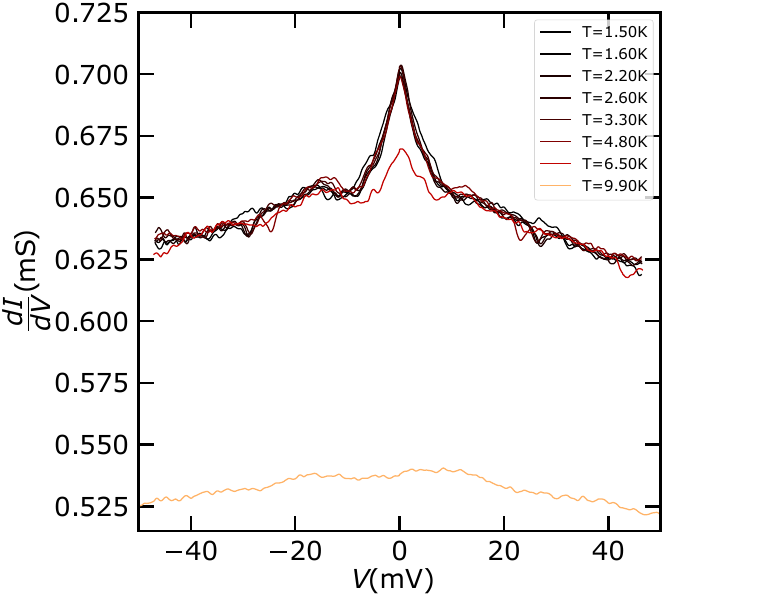}
\caption{\label{fig:Supp_Traw} Raw data of the temperature dependence of the differential conductance $dI/dV$ for different bias voltages.}
\end{figure}

\clearpage
\section{Zero bias dip in the magnetoconductance of the topological insulator ribbon}\label{sec:VII}

In several of our differential conductance measurements we observe a zero bias conductance dip for magnetic fields above 3.6$\,$T. At this magnetic field strength the superconductivity of the Nb contacts is completely suppressed since the magnetic field is above the critical field of our Nb film. However, when suppressing the superconductivity by increasing the temperature above the critical temperature $T_{\mathrm{c}}$ this zero bias resistance peak was not observed, as can be seen in the main manuscript. Stehno \textit{et. al.} also observed a zero bias dip in their measurements of a exfoliated BiSbTeSe single crystal \cite{stehno_conduction_2017}. The conductance dip has also been stable with magnetic field but unstable with increasing temperature, which is consistent with our observations. They suggested that the origin is the Altshuler–Aronov effect (electron-electron interaction) which develops in the crystal due to disorder.\\\\
In order to investigate the origin of this zero bias resistance peak in our devices, we conducted four-point measurements of our $\mathrm{(Bi_{0.15}Sb_{0.85})_2Te_3}$ nanoribbon (see Supplementary Fig.~\ref{fig:Supp_ZeroBias}(a)). 
\begin{figure}[htb]
\includegraphics[width=16cm]{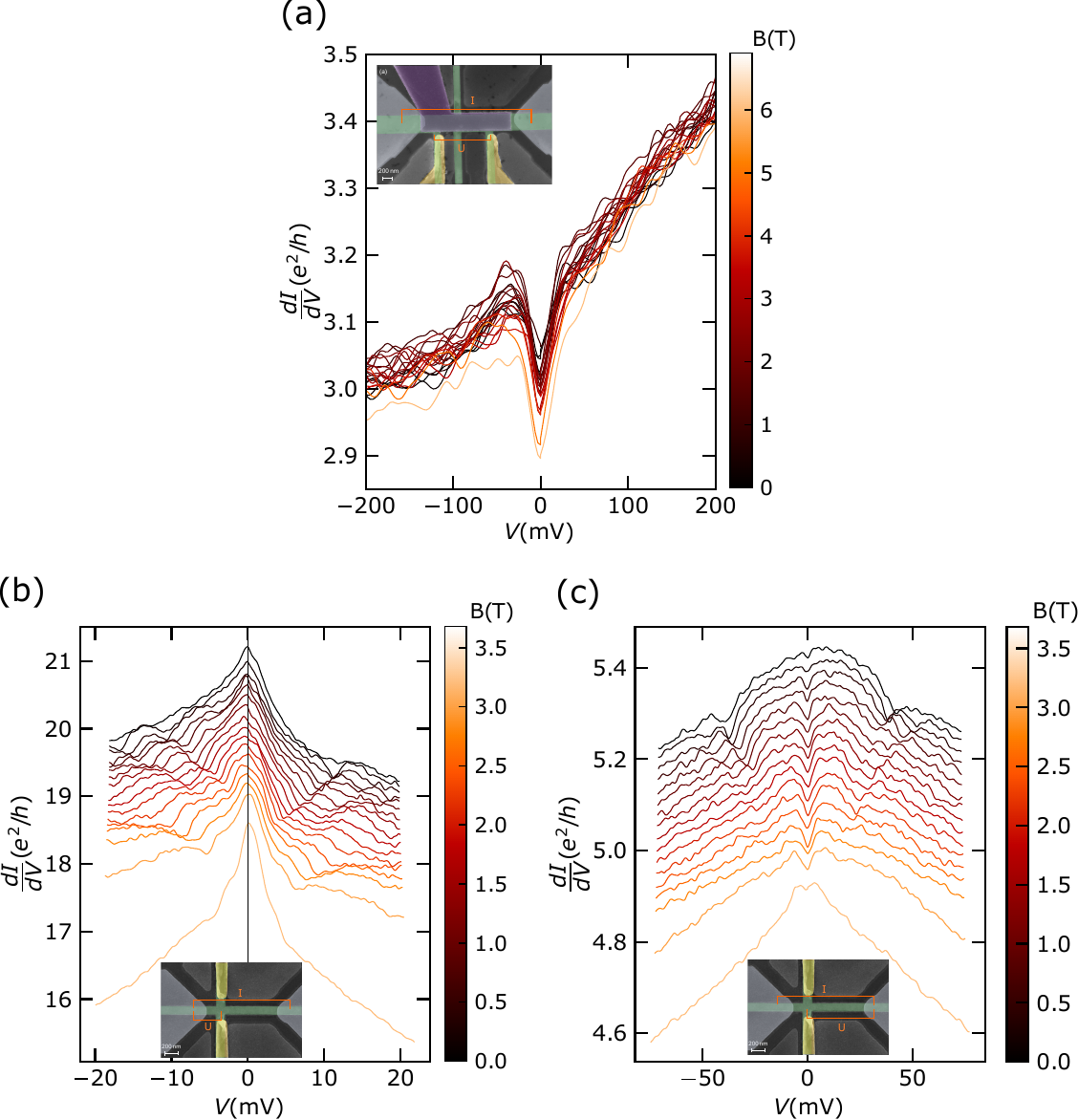}
\caption{\label{fig:Supp_ZeroBias} \textbf{(a)} Four-point differential conductance measurement as a function of voltage of a $\mathrm{(Bi_{0.15}Sb_{0.85})_2Te_3}$ ribbon for various magnetic fields. \textbf{(b)} Waterfall plot of the differential conductance of the left interface with a 200\,nm  distant close normal contact. The curves are offset by $0.1\,\mathrm{e^2/h}$. \textbf{(c)} Waterfall plot of the differential conductance of the right interface for the same device. The curves are offset by $0.02\,\mathrm{e^2/h}$. In this case with a 800\,nm distant far normal contact. Scanning electron micrographs of the corresponding devices and the electrical contacting are shown in the inset. The superconducting contacts are highlighted in grey, the normal contacts in yellow, the $\mathrm{(Bi_{0.15}Sb_{0.85})_2Te_3}$ ribbon in green.}
\end{figure}
We observed a zero bias conductance dip which is stable up to 6$\,$T, which is the highest field we applied. Due to the four-point configuration of the measurement the voltage drop only occurs over the $\mathrm{(Bi_{0.15}Sb_{0.85})_2Te_3}$ nanoribbon. Considering that temperature and a bias voltage, which is microscopically a change of the electron band structure, destroys this effect which leads to the zero bias conductance dip and magnetic field does not, this behaviour fits the expected behavior of the Altshuler-Aronov effect. 
Furthermore, this is also consistent with our observation that the dip is more pronounced when the normal contact is further apart from the interface, as can be seen in Supplementary Fig.~\ref{fig:Supp_ZeroBias}(b-c). When the normal contact is close to the interface the contribution of the $\mathrm{(Bi_{0.15}Sb_{0.85})_2Te_3}$ nanoribbon segment to the measurement signal is comparably small and no zero bias conductance dip is observed (cf. Supplementary Fig.~\ref{fig:Supp_ZeroBias}(b)). In contrast, for a large contact separation, the contribution of the nanoribbon to the measurement signal is high and the conductance dip is observed (cf. Supplementary Fig.~\ref{fig:Supp_ZeroBias}(c)).
\clearpage
\section{Transmission Electron Microscopy Data}\label{sec:VIII}

In Supplementary Fig.~\ref{fig:SuppTEM} a large scale transmission electron microscopy image of the interface region is shown. It can be seen that the interdiffusion region is limited to 2-3 quintuple layers, which roughly corresponds to 2-3\,nm. The width of the interdiffusion region agrees well with the results of Jalil \textit{et al.} \cite{jalil_engineering_2022}. Furthermore, it can be seen that the BST is brighter on the bottom half, than on the top half. This indicates that the lower part is more Bi heavy compared to the upper part, concerning its composition.
\begin{figure}[!ht]
\includegraphics[width=14cm]{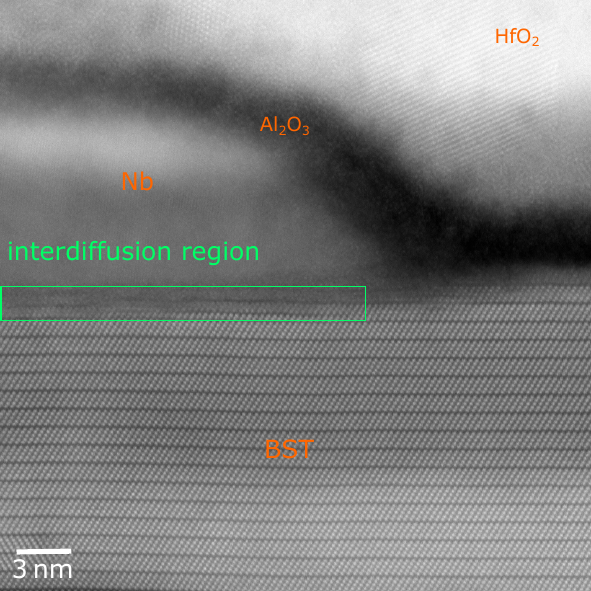}
\caption{\label{fig:SuppTEM} Large-scale high-angle annular dark-field scanning transmission electron microscopy image of the Nb/BST interface region. The interdiffusion region of 2-3\,nm is highlighted by the green box.}
\end{figure}
\clearpage

\section{Density functional theory data for bulk-insulating and $p$-type SC/TI heterostructures}\label{sec:iX}

In Supplementary Fig.~\ref{fig:SuppDFT} we compare the normal state charge density at the Fermi level for bulk-insulating (a) and $p$-type TI (b) in contact to Nb. This illustrates the increased availability of states for hybridization. The superconducting density of states (b,d) for these two situations show a broadening of the Nb coherence peak indicated by the larger FWHM, which is discussed in the main text. Note, due to the limited resolution of the energy sampling points of the DOS, the absolute high of the coherence peaks can not be compared between Fig.~\ref{fig:SuppDFT}(a) and Fig.~\ref{fig:SuppDFT}(b). 

\begin{figure}[!ht]
\includegraphics[width=0.6\textwidth]{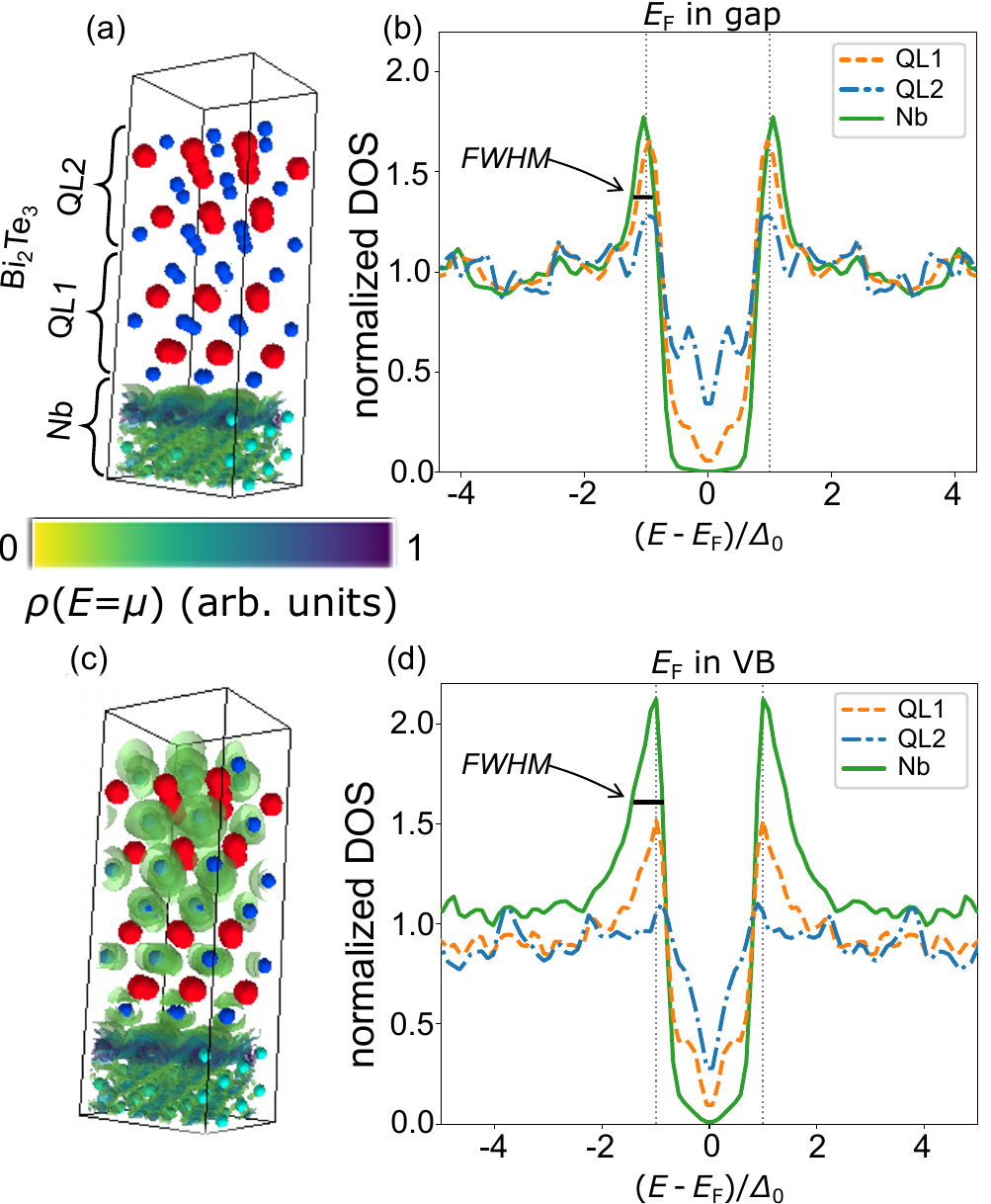}
\caption{\label{fig:SuppDFT} Comparison of bulk-insulating (a,b) and $p$-type TI (c,d) in the SC/TI heterostructure. \textbf{(a,c)} Distribution of the normal state density at the chemical potential throughout the Nb/Bi$_2$Te$_3$ heterostructure. \textbf{(b,d)} Superconducting density of states showing a broadening of the coherence peak if $\mu$ resides not in the TI bulk band gap (indicated by the larger FWHM in d compared to b).}
\end{figure}
\clearpage

\bibliography{references}
\end{document}